\documentclass[10pt,twocolumn]{article}        

\usepackage{graphicx}
\usepackage{cite}
\usepackage{amssymb,amsmath}

\begin{document}

\title{\textbf{Conservation laws for time-fractional subdiffusion \\ and diffusion-wave equations}}

\author{Stanislav Yu. Lukashchuk \\
\small {e-mail: lsu@mail.rb.ru}  \\
\small{Laboratory "Group analysis of mathematical models in natural and engineering sciences"},\\
\small{Ufa State Aviation Technical University, 12, K. Marx Str., Ufa, 450000, Russia} \\                     
}

\date{}

\maketitle

\abstract{
The concept of nonlinear self-adjointness is employed to construct the conservation laws for fractional evolution equations using its Lie point symmetries. The approach is demonstrated on subdiffusion and diffusion-wave equations with the Riemann-Li\-ou\-vil\-le and Caputo time-fractional derivatives. It is shown that  these equations are nonlinearly self-adjoint and therefore desired conservation laws can be obtained using appropriate formal Lagrangians. Fractional generalizations of the Noether operators are also proposed for the equations with the Riemann-Liouville and Caputo time-fractional derivatives of order $\alpha \in (0,2)$. Using these operators and formal Lagrangians, new conserved vectors have been constructed for the linear and nonlinear fractional subdiffusion and diffusion-wave equations corresponding to its Lie point symmetries.}

%\keywords{Time-fractional diffusion equation \and Conservation law \and Nonlinear self-adjointness \and Symmetry}

\section{Introduction}
\label{intro}

A problem of constructing conservation laws for the fractional differential equations is considered. It is well-known that if differential equation is an Euler-Lagrange equation, then conservation laws can be found using Noether's theorem by variational Lie point symmetries of this equation. Recall that an Euler-Lagrange differential equation is obtained from the variational principle of least action by minimization of variational integral with a Lagrangian function as integrand. In 1996, Riewe \cite {R_96} introduce a Lagrangian depending on fractional derivatives. During last two decades, many fractional generalizations of the Euler-Lagrange equations with different types of fractional derivatives have been derived \cite{R_96, Agr_JMAA, Agr_CMA, AMB, HB, LT}. Using these results, fractional generalizations of Noether's theorem have been proposed \cite{FT_07, AKPS, Mal, OMT, BCG, LZ}. Also, several fractional conservation laws have been obtained for equations and systems that possess fractional Lagrangians \cite{AKPS, FT_08, ZCF}. Nevertheless, most physically justified fractional differential equations such as fractional diffusion and transport equations, fractional kinetic equations, fractional relaxation equations (see, e.g., \cite{MK, Hil, KRS, Mai, KLM, BDST, US} and references therein) are not the Euler-Lagrange equations and do not have fractional La\-grang\-ians. Therefore, the conservation laws can not be obtained for these equations by Noether's theorem, and a new approach has to be found. 

Recently for the integer-order differential equations, Ibragimov  \cite{Ibr_MMAA, Ibr_JPA, Ibr_ALGA, Ibr_RMS} introduced the general concept of nonlinear self-adjointness which is based on the notion of \textit{formal Lagrangian}. He proves that conservation laws can be associated with the symmetries of nonlinearly self-adjoint differential equations and its systems. He also shows that the constructive algorithm proposed earlier \cite{Ibr_83, Ibr_B} is applicable for these equations. The components of conserved vectors are obtained in this algorithm by acting the so-called Noether operators to a Lagrangian function. Using this approach, the conservation laws for different integer-order differential equations and its systems having only formal Lagrangians have been constructed by their symmetries \cite{Ibr_JPA, Ibr_ALGA, Ibr_RMS, Ibr_MMNP, GBR, BDI, AIL, BIZY, G}.  

In this paper, it is shown that the concept of nonlinear self-adjointness is applicable to the fractional differential equations that do not have fractional Lagrangians in classical sense, and that this concept can be used to construct conservation laws for such equations. The time-fractional diffusion equations are used to demonstrate the approach. Cases of linear and nonlinear subdiffusion and diffusion-wave equations with the Riemann-Li\-ou\-vil\-le and Caputo time-fractional derivatives are considered separately. The nonlinear self-ad\-joint\-ness of these equations are established, and corresponding formal fractional Lagrangians have been found. Fractional generalizations of the Noether operators applicable to these equations are also presented in the explicit forms. Using these fractional Noether operators and formal Lagrangians, the conserved vectors corresponding to the group of point transformations admitted by the time-fractional subdiffusion and diffusion-wave equations have been constructed.

%-----------------------------------------------------
\section{Description of the approach}
\label{sec:1}
%-----------------------------------------------------

\subsection{Time-fractional diffusion equations}

Let us consider a nonlinear time-fractional diffusion equation (TFDE)
\begin{equation}
\label{TFDE}
\mathcal{D}^{\alpha}_t u=(k(u)u_x)_x,  \ \ \ \alpha \in (0,2).
\end{equation}
Here $u$ is a function of independent variables $t \in (0,T]$ $(T \le \infty)$ and $x \in \Omega \subset {\mathbb R}$, $\mathcal{D}^{\alpha}_t u$ is a fractional derivative of function $u$ with respect to $t$ of order $\alpha$. 

In this paper, two different types of fractional derivatives will be used as $\mathcal{D}^{\alpha}_t u$ in Eq.~(\ref{TFDE}): one is the Riemann-Liouville left-sided time-fractional derivative ${}^{}_0D^{\alpha}_t u$, and the other is the Caputo left-sided time-fractional derivative ${}^{C}_0D^{\alpha}_t u$. These fractional derivatives are defined by  
%\begin{align*}
$$
{}^{}_0D^{\alpha}_t u = D_t^n \left({}^{}_0I^{n-\alpha}_t u\right), \ \ \ {}^{C}_0D^{\alpha}_t u = {}^{}_0I^{n-\alpha}_t \left(D_t^n u\right)
$$
(see, e.g., \cite{BDST, US, SKM, KST, U}). Here $D_t$ is the operator of differentiation with respect to $t$, $n=[\alpha]+1$, and ${}^{}_0I^{n-\alpha}_t u$ is the left-sided time-fractional integral of order $n-\alpha$ defined by
$$
({}^{}_0I^{n-\alpha}_t u)(t,x)=\frac{1}{\Gamma(n-\alpha)} \int_0^t \frac{u(\tau,x)}{(t-\tau)^{1-n+\alpha}} d\tau,
$$
where $\Gamma(z)$ is the Gamma function.

Eq.~(\ref{TFDE}) is known as subdiffusion equation for $\alpha \in (0,1)$ and as diffusion-wave equation for $\alpha \in (1,2)$. 

It is necessary to note that TFDEs with the Riemann-Liouville and with the Caputo fractional derivatives demonstrate different behavior. For $\alpha=1$ Eq.~(\ref{TFDE}) with the Riemann-Liouville fractional derivative coincides with the classical diffusion equation, and for $\alpha=2$ it coincides with the classical wave equation. For the Caputo fractional derivative, Eq.~(\ref{TFDE}) coincides with the classical diffusion and wave equations only in the left-side limits $\alpha \rightarrow 1_{-}$ and $\alpha \rightarrow 2_{-}$. Thus, the diffusion-wave equation of order $\alpha \in (1,2)$ with the Riemann-Liouville fractional derivative provides the continuous passage from the diffusion equation to wave equation. The diffusion-wave equation with the Caputo fractional derivative does not possess this property. Therefore, equations with the Riemann-Liouville fractional derivatives seem to be more preferable as mathematical models of anomalous diffusion processes. 

Nevertheless, TFDEs with the Caputo fractional de\-ri\-va\-tives are more frequently used in practice because they have more natural initial conditions. The solutions of TFDEs with the Riemann-Liouville time-fractional derivatives usually have a singularity at the initial time point $t=0$. The physical meaning of this singularity generally is not clear. However, it is necessary to note that TFDEs usually derived using power-law assymptotics and therefore these equations are not valid in the neighborhood of the initial time point $t=0$ (see, e.g., \cite{LSY} for detailed discussion). Also contrary to the integer-order evolution equations, time-fractional differential equations do not admit translation in time. Hence, TFDEs with the Riemann-Liouville time-frac\-ti\-o\-nal de\-ri\-va\-tives can be used for valid modeling of anomalous diffusion processes only for time $t>t_0$, where $t_0$ is not small enough. 

So, both the Riemann-Liouville and Caputo time-fractional derivatives can be used in practice for modeling anomalous diffusion procesess. Therefore, in this paper Eq.~(\ref{TFDE}) will be considered for both this types of fractional derivatives. 

The symmetry properties of Eq.~(\ref{TFDE}) have been investigated in \cite{GKL_PS}, and Lie point symmetries for this equation with the Riemann-Liouville and Caputo time-fractional derivatives have been obtained there. These symmetries will be used in this paper to construct the conservation laws for Eq.~(\ref{TFDE}).

\subsection{Conservation laws and nonlinear self-adjointness}

In this paper, a conservation law for Eq.~(\ref{TFDE}) is defined in the same manner as it defines for the classical diffusion and wave equations. Namely, a vector field $C=(C^t,C^x)$ where $C^t=C^t(t,x,u,\ldots)$, $C^x=C^x(t,x,u,\ldots)$ is called a \textit{conserved vector} for Eq.~(\ref{TFDE}) if it satisfies the conservation equation 
\begin{equation}
\label{CL}
D_t C^t+ D_x C^x =0
\end{equation}
on all solutions of Eq.~(\ref{TFDE}). Eq.~(\ref{CL}) is called a \textit{conservation law} for Eq.~(\ref{TFDE}).

A conserved vector is called a \textit{trivial conserved vector} for Eq.~(\ref{TFDE}) if its components $C^t$ and $C^x$ vanish on the solution of this equation.

Note that Eq.~(\ref{TFDE}) with the Riemann-Liouville fractional derivative can be rewritten in the form of conservation law form (\ref{CL}) with 
\begin{equation}
\label{TCVRL}
C^t=D^{n-1}_t\left({}^{}_0I^{n-\alpha}_t u\right), \ \ \ C^x=-k(u)u_x, \ \ \ n=1,2.
\end{equation}
It is important to point out that in (\ref{TCVRL}), the order $n-\alpha$ of fractional integral is the same as the one used in Eq.~(\ref{TFDE}).

In the case of the Caputo fractional derivative, Eq.~(\ref{TFDE}) can also be rewritten in the form of conservation law (\ref{CL}) with
\begin{equation}
\label{TCVC}
C^t={}^{}_0I^{n+1-\alpha}_t \left(D^n_t u\right), \ \ \ C^x=-k(u)u_x, \ \ \ n=1,2.
\end{equation}
Contrary to the previous case, the order of fractional integral in (\ref{TCVC}) has been increased by one. In other words, the coordinate $C^t$ now depends on a new integral variable.

As well as the classical diffusion equation, Eq.~(\ref{TFDE}) is not an Euler-Lagrange equation in classical sense. This means that Eq.~(\ref{TFDE}) can not be derived from the variational principle of least action with a Lagrangian depending on the variables $t$, $x$, $u$ and any integer-order and/or fractional-order integrals and derivatives of $u$. So, Eq.~(\ref{TFDE}) does not have a Lagrangian in classical sense. 

Nevertheless in accordance with the concept of nonlinear self-adjointness \cite{Ibr_JPA}, a formal Lagrangian for this equation can be introduced as
\begin{equation}
\label{FL}
\mathcal{L}=v(t,x) \left[ \mathcal{D}^{\alpha}_t u-k'(u) u^2_x -k(u) u_{xx} \right],
\end{equation}
where $v$ is a new dependent variable. In view of this formal Lagrangian, an action integral is defined by
\begin{equation}
\label{Act}
\int_0^T \int_{\Omega} \mathcal{L}(t,x,u,v,\mathcal{D}^{\alpha}_t u, u_x, u_{xx}) dx dt. 
\end{equation}

Assume that variable $v$ in the action (\ref{Act}) is not varied. Then using fractional variational approach developed by Agrawal \cite{Agr_JMAA}, one can find the Euler-La\-gran\-ge operator with respect to $u$ corresponding to the action (\ref{Act}) as
\begin{equation}
\label{ELO}
\frac{\delta}{\delta u} = \frac{\partial}{\partial u} + \left(\mathcal{D}^{\alpha}_t\right)^{*} \frac{\partial}{\partial \mathcal{D}^{\alpha}_t u} - D_x \frac{\partial}{\partial u_x} +D^2_x \frac{\partial}{\partial u_{xx}}.
\end{equation}
Here $\left(\mathcal{D}^{\alpha}_t\right)^{*}$ is the adjoint operator of $\mathcal{D}^{\alpha}_t$. For the Rie\-mann-Liouville and Caputo fractional differential operators, the corresponding adjoint operators have the form  
$$
\left({}^{}_0D^{\alpha}_t \right)^{*} = (-1)^n {}^{}_tI^{n-\alpha}_T \left(D^n_t\right) \equiv {}^{C}_t{D}^{\alpha}_T,
$$
$$
\left({}^{C}_0{D}^{\alpha}_t \right)^{*} = (-1)^n D^n_t \left({}^{}_tI^{n-\alpha}_T \right)\equiv {}^{}_tD^{\alpha}_T.
$$
Here ${}^{}_t{I}^{n-\alpha}_T$ is the right-sided operator of fractional integration of order $n-\alpha$ defined by
$$
\left({}^{}_tI^{n-\alpha}_T f\right)(t,x)=\frac{1}{\Gamma(n-\alpha)}\int_t^T \frac{f(\tau,x)}{(\tau-t)^{\alpha+1-n}} d \tau,
$$
${}^{}_tD^{\alpha}_T$ and ${}^{C}_t{D}^{\alpha}_T$ are the right-sided Riemann-Liouville and Caputo operators of fractional differentiation of order $\alpha$, respectively. 

Similarly to the case of integer-order nonlinear differential equations \cite{Ibr_JPA, Ibr_ALGA, Ibr_RMS}, the \textit{adjoint equation} to the nonlinear TFDE (\ref{TFDE}) can be defined as  Euler-Lagrange equation
\begin{equation}
\label{ELE}
\frac{\delta \mathcal{L}}{\delta u}=0,
\end{equation}
where $\mathcal{L}$ is the formal Lagrangian (\ref{FL}) and $\frac{\delta}{\delta u}$ is the Euler-Lagrange operator (\ref{ELO}). After calculations, Eq.~(\ref{ELE}) takes the form
\begin{equation}
\label{AE}
\left( \mathcal{D}^{\alpha}_t \right)^{*} v - k(u) v_{xx} =0, \ \ \ n=[\alpha]+1, \ \ \ \alpha \in (0,2).
\end{equation} 

Now, definition of nonlinear self-adjointness (see, e.g., definition 2 from \cite{Ibr_JPA}) can be extended to the time-fractional diffusion equations. Namely, Eq.~(\ref{TFDE}) will be called a \textit{nonlinearly self-adjoint} if the adjoint equation (\ref{AE}) is satisfied for all solutions $u$ of the Eq.~(\ref{TFDE}) upon substitution $v=\varphi(t,x,u)$ such that $\varphi(t,x,u) \ne 0$. Specific types of this substitution are presented in the following sections. After this substitution, the formal Lagrangian (\ref{FL}) can be used as usual classical Lagrangian for constructing conservation laws for Eq.~(\ref{TFDE}) using fractional generalizations of Noether's theorem.

\subsection{Fractional Noether operators}

Nevertheless, it is unwieldy to construct the conserved vectors by direct using of Noether's theorem (see, e.g., \cite{AKPS, ZCF}). More convenient approach for integer-order differential equations was proposed in \cite{Ibr_83}. In this approach, the components of conserved vector are obtained by acting the so-called Noether operators to the Lagrangian. These Noether operators can be found from the \textit{fundamental operator identity}, also known as the Noether identity. For the considered case of two independent variables $t$, $x$, and one dependent variable $u(t,x)$, this fundamental identity can be written as
\begin{equation}
\label{NI}
\tilde{X}+D_t(\xi^0) \mathcal{I}+D_x(\xi^1) \mathcal{I} = W\frac{\delta}{\delta u}+D_t \mathcal{N}^t+D_x \mathcal{N}^x.
\end{equation}
Here $\mathcal{I}$ is the identity operator, $\frac{\delta}{\delta u}$ is the Euler-La\-gran\-ge operator, $\mathcal{N}^t$ and $\mathcal{N}^x$ are the Noether operators, 
$\tilde{X}$ is an appropriate prolongation for the Lie point group generator 
\begin{equation}
\label{X}
X=\xi^0(t,x,u)\frac{\partial}{\partial t}+\xi^1(t,x,u)\frac{\partial}{\partial x}+\eta(t,x,u)\frac{\partial}{\partial u}
\end{equation}
to all derivatives (integer and/or fractional order) of dependent variable $u(t,x)$ are contained in considered equation, and $W=\eta-\xi^0 u_t-\xi^1 u_x$.

The prolongation of a group of point transformations acting on a space with fractional variables have been discussed in \cite{GKL_PS, GKL_VU, GKL_UMJ}, and corresponding prolongation formulae have been presented there. The prolongation of the generator (\ref{X}) for Eq.~(\ref{TFDE}) has the form
\begin{multline}
\label{X_prol}
\tilde{X}=\xi^0\frac{\partial}{\partial t}+\xi^1\frac{\partial}{\partial x}+\eta\frac{\partial}{\partial u} \\
+\zeta^0_{\alpha}\frac{\partial}{\partial (\mathcal{D}^{\alpha}_t u)}+
\zeta^1_1 \frac{\partial}{\partial u_x}+\zeta^1_2 \frac{\partial}{\partial u_{xx}},
\end{multline}
where $\zeta^0_{\alpha}, \ \zeta^1_1, \ \zeta^1_2$ are given by the prolongation formulae
\begin{align*}
\zeta^0_{\alpha} & =\mathcal{D}^{\alpha}_t(W)+\xi^0 D_t(\mathcal{D}^{\alpha}_t u)+\xi^1 D_x (\mathcal{D}^{\alpha}_t u), \\
\zeta^1_1 			 & =D_x(W)+\xi^0 u_{tx}+\xi^1 u_{xx}, \\ 
\zeta^1_2        & =D^2_x(W)+\xi^0 u_{txx}+\xi^1 u_{xxx}.
\end{align*}

For a given operators (\ref{ELO}) and (\ref{X_prol}), one can verify that the equality (\ref{NI}) is fulfilled if the Noether operators are defined as follows. For the case when the Riemann-Liouville time-fractional derivative is used in Eq.~(\ref{TFDE}), the operator $\mathcal{N}^t$ is given by
\begin{multline}
\label{NOTRL}
\mathcal{N}^t=\xi^0 \mathcal{I}+\sum_{k=0}^{n-1} (-1)^k {}^{}_0D^{\alpha-1-k}_t (W) D^k_t \frac{\partial}{\partial \left({}^{}_0D^{\alpha}_t u\right)} \\
															 -(-1)^{n} J\left(W,D^n_t \frac{\partial}{\partial \left({}^{}_0D^{\alpha}_t u\right)} \right). 
\end{multline}
For the another case when the Caputo time-fractional derivative is used in Eq.~(\ref{TFDE}), this operator takes the form 
\begin{multline}
\label{NOTC}
\mathcal{N}^t=\xi^0 \mathcal{I}+\sum_{k=0}^{n-1} D^{k}_t (W) {}^{}_tD^{\alpha-1-k}_T \frac{\partial}{\partial \left({}^{C}_0D^{\alpha}_t u\right)} \\
															 -J\left(D^n_t(W),\frac{\partial}{\partial \left({}^{C}_0D^{\alpha}_t u\right)} \right). 
\end{multline}
The operator $\mathcal{N}^x$ in both cases is defined by
\begin{equation}
\label{NOX}
\mathcal{N}^x=\xi^1 \mathcal{I}+W\left(\frac{\partial}{\partial u_x}-D_x \frac{\partial}{\partial u_{xx}} \right)+D_x(W)\frac{\partial}{\partial u_{xx}}.
\end{equation}
In (\ref{NOTRL}) and (\ref{NOTC}), $J$ is the integral
\begin{equation}
\label{J}
J(f,g)= \frac{1}{\Gamma(n-\alpha)}\int_0^t \int_t^T \frac{f(\tau,x)g(\mu,x)}{(\mu-\tau)^{\alpha+1-n}} d \mu d \tau.
\end{equation}
This integral has a property 
$$
D_t J(f,g) = f \, {}^{}_tI^{n-\alpha}_T g-g \, {}^{}_0I^{n-\alpha}_t f.
$$
%\end{equation}

Note that in the specific case of one dependent variable $t$ and $\alpha \in (0,1)$, integral (\ref{J}) coincides with the third integral in the fractional conservation law presented in \cite{AKPS}. 

To the best of the author's knowledge, operators $\mathcal{N}^t$ defined by (\ref{NOTRL}) and (\ref{NOTC}) are for the first time presented in this paper. It can be proved that in the limiting case of integer $\alpha$ these operators coincide with the known integer-order Noether operators presented in \cite{Ibr_83} (see also \cite{Ibr_MMAA, Ibr_ALGA, Ibr_B}).

Now assume that Eq.~(\ref{TFDE}) is nonlinearly self-adjoint. This means that a function $v=\varphi(t,x,u)$ exists such that Eq.~(\ref{AE}) is satisfied for any solution of Eq.~(\ref{TFDE}). Then the explicit formulae for the components of conserved vectors associated with symmetries of Eq.~(\ref{TFDE}) can be established. 

We act on the formal Lagrangian (\ref{FL}) by both sides of the Noether identity (\ref{NI}). For any generator $X$ admitted by Eq.~(\ref{TFDE}) and any solution of this equation, the left-hand side of this equality is equal to zero:
$$
\left. \left( \tilde{X}\mathcal{L}+D_t(\xi^0) \mathcal{L} +D_x(\xi^1) \mathcal{L} \right) \right|_{(\ref{TFDE})} = 0.
$$
Because for nonlinearly self-adjoint equations the Euler-Lagrange equation (\ref{ELE}) is valid, the right-hand side of the equality under consideration leads to the conservation law
\begin{equation}
\label{CLN}
D_t (\mathcal{N}^t \mathcal{L})+D_x (\mathcal{N}^x \mathcal{L}) =0,
\end{equation}
where operator $\mathcal{N}^t$ is defined by (\ref{NOTRL}) or (\ref{NOTC}), and operator $\mathcal{N}^x$ is defined by (\ref{NOX}).

From the comparison of (\ref{CL}) and (\ref{CLN}) it is easy to conclude that, any Lie point symmetry of Eq.~(\ref{TFDE}) gives the conserved vector for this equation with components defined by the explicit formulae
\begin{equation}
\label{Comp_CV}
C^t=\mathcal{N}^t(\mathcal{L}), \ \ \ C^x=\mathcal{N}^x(\mathcal{L}).
\end{equation}

In the following sections, it is proved that Eq.~(\ref{TFDE}) is nonlinearly self-adjoint and conserved vectors associated with different symmetries of this equation are constructed.

%-----------------------------------------------------
\section{Conservation laws for linear TFDE}
\label{sec:2}
%-----------------------------------------------------

\subsection{Nonlinear self-adjointness of linear TFDE}

At first, let us consider a simple case when the diffusion coefficient $k$ in Eq.~(\ref{TFDE}) does not depend on function $u$, i.e. $k=const$. Then Eq.~(\ref{TFDE}) is linear. With no loss of generality, one can set $k=1$. It was shown in \cite{GKL_PS} that, for both considered types of fractional derivatives and all $\alpha \in (0,2)$, the corresponding Lie algebra of point symmetries is infinite and is spanned by generators
\begin{align}
\label{X_k_1}
X_1 & = \displaystyle\frac{\partial}{\partial x},  & & X_2 = 2t\frac{\partial}{\partial t}+\alpha x \frac{\partial}{\partial x}, \nonumber \\
X_3 & = u\displaystyle\frac{\partial}{\partial u}, & & X_{\infty} = h\frac{\partial}{\partial u}.
\end{align}
Here $h=h(t,x)$ is an arbitrary solution of the equation $\mathcal{D}^{\alpha}_t h = h_{xx}$.

In the considered linear case, the adjoint equation (\ref{AE}) takes the form 
\begin{equation}
\label{AE_1}
\left( \mathcal{D}^{\alpha}_t \right)^{*} v = v_{xx}.
\end{equation}
It can be seen that this equation is also linear and does not contain function $u$. 

Let $v=\varphi(t,x) \neq 0$ is an arbitrary non-trivial solution of this adjoint equation. Because (\ref{AE_1}) is satisfied upon the substitution $v=\varphi(t,x)$ for all $u(t,x)$ then in accordance with the definition of nonlinear self-adjointness given in the previous section, the linear equation (\ref{TFDE}) is nonlinearly self-adjoint with a such function $\varphi(t,x)$. Note, that the adjoint equation (\ref{AE_1}) has nontrivial solutions. For example, particular nontrivial solutions of Eq.~(\ref{AE_1}) are $v(t,x)=ct^{\alpha-1}x$ for $\mathcal{D}^{\alpha}_t = {}^{}_0D^{\alpha}_t$, and $v(t,x)=ctx$ for $\mathcal{D}^{\alpha}_t = {}^{C}_0D^{\alpha}_t$ (here $c$ is an arbitrary constant).

The formal Lagrangian (\ref{FL}) for linear Eq.~(\ref{TFDE}) has the form 
\begin{equation}
\label{FL_lin}
\mathcal{L}=\varphi(t,x) \left[ \mathcal{D}^{\alpha}_t u-u_{xx} \right].
\end{equation} 
Using this Lagrangian, one can find the conserved vectors for linear equation (\ref{TFDE}) corresponding to the symmetries (\ref{X_k_1}) by the formulae (\ref{Comp_CV}) with the Noether operators defined by (\ref{NOTRL}) -- (\ref{NOX}).

%%%%%%%%%%%%%%%%%%%%%
\subsection{Conservation laws for TFDE with the Riemann-Liouville fractional derivative}
%%%%%%%%%%%%%%%%%%%%%

Calculations by (\ref{Comp_CV}) give the following results for the components of conserved vectors for Eq.~(\ref{TFDE}) with the Riemann-Liouville fractional derivative.

For the subdiffusion equation when $\alpha \in (0,1)$ the components of conserved vectors are given by
\begin{align*}
&C^t_i=\varphi \, {}^{}_0I^{1-\alpha}_t (W_i) + J(W_i,\varphi_t), \nonumber \\ 
&C^x_i=\varphi_x W_i  - \varphi W_{ix}. \ \ \,
\end{align*}
Here subscript $i$ coincides with the number of appropriate symmetry from (\ref{X_k_1}) ($i=1,2,3$ and $\infty$), and functions $W_i$ have the form 
\begin{align}
\label{WX}
& W_1=u_x, & & W_2=2 t u_t+\alpha x u_x, \nonumber \\ 
& W_3=u, & & W_{\infty}=h.
\end{align}

In the same way for the diffusion-wave equation when $\alpha \in (1,2)$, the components of conserved vectors are given by
\begin{align*}
& C^t_i=\varphi \, {}^{}_0D^{\alpha-1}_t (W_i) -\varphi_t \, {}^{}_0I^{2-\alpha}_t (W_i) -J(W_i,\varphi_{tt}), \nonumber \\ 
& C^x_i=\varphi_x W_i  - \varphi W_{ix}, \ \ \ i=1,2,3,\infty.
\end{align*}
Here functions $W_i$ are defined by (\ref{WX}).

Because operators $D_x$ and ${}^{}_0D^{\alpha}_t$ commute with each other then, the conserved vectors corresponding to generators $X_1$ and $X_2$ can be rewritten in another form. Hereafter $w=2t\varphi_t+\alpha x \varphi_x$.

\textbf{Case 1.} \textit{Subdiffusion equation} ($0 < \alpha < 1$).
\begin{equation*} 
\begin{array}{ll}
X_1: 				& C^t_1=\varphi_x \, {}^{}_0I^{1-\alpha}_t u +J(u,\varphi_{tx}), \\ \vspace{4pt}
     				& C^x_1=-\varphi_x u_x + \varphi_{xx} u; \\ 
X_2: 				& C^t_2=w \, {}^{}_0I^{1-\alpha}_t u - 2t \left[ \varphi_{t} \, {}^{}_0I^{1-\alpha}_t u - u \, {}^{}_tI^{1-\alpha}_T \varphi_{t} \right] \\ 
						& \ \ \ \ \ \ \ +2J(tu_t-(\alpha-1)u,\varphi_t)-\alpha x J(u,\varphi_{tx}), \\ 
		 				& C^x_2=-w u_x + w_x u.	\\ 		 				
\end{array}
\end{equation*}

\textbf{Case 2.} \textit{Diffusion-wave equation} ($1 < \alpha < 2$).
\begin{equation*}
\begin{array}{ll}
X_1: 				& C^t_1=\varphi_x \, {}^{}_0D^{\alpha-1}_t u -\varphi_{tx} \, {}^{}_0I^{2-\alpha}_t u-J(u,\varphi_{ttx}), \\  \vspace{4pt}
     				& C^x_1=-\varphi_x u_x +\varphi_{xx} u; \\ 
X_2: 				& C^t_2=w \, {}^{}_0D^{\alpha-1}_t u - w_t \, {}^{}_0I^{2-\alpha}_t u \\
						& \ \ \ \ \ \ \	+2t \left[ \varphi_{tt} \, {}^{}_0I^{2-\alpha}_t u - u \, {}^{}_tI^{2-\alpha}_T \varphi_{tt} \right] \\ 
						& \ \ \ \ \ \ \ +2J(tu_t-(\alpha-1)u,\varphi_{tt})-\alpha x J(u,\varphi_{ttx}), \\ 
		 				& C^x_2=-w u_x + w_x u.	\\ 		 				
\end{array}
\end{equation*}

%%%%%%%%%%%%%%%%%%%%%
\subsection{Conservation laws for TFDE with the Caputo fractional derivative}
%%%%%%%%%%%%%%%%%%%%%

Using (\ref{Comp_CV}) with the formal Lagrangian (\ref{FL_lin}) and the Noether operators (\ref{NOTC}), (\ref{NOX}), the conserved vectors corresponding to the symmetries (\ref{X_k_1}) for linear Eq.~(\ref{TFDE}) with the Caputo fractional derivative have been found. 

For the subdiffusion equation when $\alpha \in (0,1)$ one can find
\begin{align*}
&C^t_i=W_i \, {}^{}_tI^{1-\alpha}_T (\varphi) - J(W_{it},\varphi), \nonumber \\ 
&C^x_i=\varphi_x W_i  - \varphi W_{ix}, \ \ \ i=1,2,3,\infty.
\end{align*}
For the diffusion-wave equation when $\alpha \in (1,2)$ the components of conserved vectors can be written as
\begin{align*}
& C^t_i=W_i \, {}^{}_tD^{\alpha-1}_T \varphi + W_{it} \, {}^{}_tI^{2-\alpha}_T \varphi - J(W_{itt},\varphi), \nonumber \\
& C^x_i=\varphi_x W_i  - \varphi W_{ix}, \ \ \ i=1,2,3,\infty.
\end{align*}
As previously, functions $W_i$ are defined by (\ref{WX}). 

Similarly to the case of the Riemann-Liouville fractional derivative, the conserved vectors corresponding to generators $X_1$ and $X_2$ can be presented in another form.

\textbf{Case 1.} \textit{Subdiffusion equation} ($0 < \alpha < 1$).
\begin{equation*}
\begin{array}{ll}
X_1: 				& C^t_1= u \, {}^{}_tI^{1-\alpha}_T \varphi_x - J(u_t,\varphi_{x}), \\ \vspace{4pt}
     				& C^x_1=-\varphi_x u_x + \varphi_{xx} u; \\ 
X_2: 				& C^t_2=-\displaystyle\frac{2T}{\Gamma(1-\alpha)}\frac{u \varphi(T,x)}{(T-t)^{\alpha}}  \\
					  & \ \ \ \ \ \ \ +u \, {}^{}_tI^{1-\alpha}_T w - 2t \left[ u_{t} \, {}^{}_tI^{1-\alpha}_T \varphi - \varphi \, {}^{}_0I^{1-\alpha}_t u_t \right] \\ 
						& \ \ \ \ \ \ \ + 2J(tu_{tt}-(\alpha-2)u_t,\varphi)-\alpha x J(u_t,\varphi_x), \\ 
		 				& C^x_2=-w u_x + w_x u.	\\ 
\end{array}
\end{equation*}

\textbf{Case 2.} \textit{Diffusion-wave equation} ($1 < \alpha < 2$).
\begin{equation*}
\begin{array}{ll}
X_1: 				& C^t_1=u \, {}^{}_tD^{\alpha-1}_T \varphi_x +u_t \, {}^{}_tI^{2-\alpha}_T \varphi_x - J(u_{tt},\varphi_x), \\  \vspace{4pt}
     				& C^x_1=-\varphi_x u_x +\varphi_{xx} u; \\ 
X_2: 				& C^t_2=-\displaystyle\frac{2T}{\Gamma(1-\alpha)}\frac{u \varphi(T,x)}{(T-t)^{\alpha}}-\displaystyle\frac{2T}{\Gamma(2-\alpha)}\frac{u_t  
                  \varphi(T,x)}{(T-t)^{\alpha-1}}  \\
					  & \ \ \ \ \ \ \ + u_t \, {}^{}_tI^{2-\alpha}_T w + u \, {}^{}_tD^{\alpha-1}_T w \\
					  & \ \ \ \ \ \ \ - 2t \left[ u_{tt} \, {}^{}_tI^{2-\alpha}_T \varphi - \varphi \, {}^{}_0I^{2-\alpha}_t u_{tt} \right] \\ 
						& \ \ \ \ \ \ \ + 2J(tu_{ttt}-(\alpha-3)u_{tt},\varphi)-\alpha x J(u_{tt},\varphi_x), \\ 
		 				& C^x_2=-w u_x + w_x u.	\\  				
\end{array}
\end{equation*}

%-----------------------------------------------------
\section{Conservation laws for nonlinear TFDE}
\label{sec:3}
%-----------------------------------------------------

\subsection{Symmetries and nonlinear self-adjointness of nonlinear TFDE}

Now let us consider a general case when the diffusion coefficient $k(u) \neq const$. It was shown in \cite{GKL_PS} that, Eq.~(\ref{TFDE}) in both cases of the Riemann-Liouville and Caputo fractional derivatives for arbitrary $k(u)$ and $\alpha \in (0,2)$ has two-dimensional Lie algebra of point symmetries spanned by generators
\begin{equation}
\label{X_Abtr_k}
X_1 = \frac{\partial}{\partial x}, \ \ \ X_2=2t\frac{\partial}{\partial t}+\alpha x \frac{\partial}{\partial x}.
\end{equation}

This algebra extends in some special cases of $k(u)$. 

If $k(u)=u^\beta$ ($\beta \neq 0$), then Eq.~(\ref{TFDE}) has an additional symmetry 
\begin{equation}
\label{X_k_ub}
X^{(1)}_3=\beta x \frac{\partial}{\partial x} + 2u\frac{\partial}{\partial u}
\end{equation}
for any $\alpha \in (0,2)$. 
If $\beta=-4/3$, i.e. $k(u)=u^{-4/3}$, there is an additional extension
\begin{equation}
\label{X_k_u_43}
X^{(1)}_4=x^2 \frac{\partial}{\partial x} - 3xu\frac{\partial}{\partial u}.
\end{equation}

Eq.~(\ref{TFDE}) with the Riemann-Liouville fractional derivative also admits generator
\begin{equation}
\label{X_k_u_a}
X^{(2)}_4=t^2 \frac{\partial}{\partial t}+(\alpha-1)tu\frac{\partial}{\partial u}
\end{equation}
for $k(u)=u^{\beta}$ with $\beta=-2\alpha/(\alpha-1)$, $\alpha \in (0,2)$.

Finally, if $k(u)=e^u$, Eq.~(\ref{TFDE}) with the Caputo fractional derivative of order $\alpha \in (0,2)$ has a symmetry
\begin{equation}
\label{X_k_e_u}
X^{(2)}_3=x \frac{\partial}{\partial x}+2\frac{\partial}{\partial u}.
\end{equation}

Now, the adjoint equation (\ref{AE}) depends on the function $u$. Nevertheless, there are specific solutions of this equation that do not depend on this function. 

If Eq.~(\ref{TFDE}) with the Riemann-Liouville fractional de\-ri\-va\-tive is considered, then the right-sided Caputo fractional derivative is used in the corresponding adjoint equation (\ref{AE}). This adjoint equation for the subdiffusion regime $\alpha \in (0,1)$ has a solution 
\begin{equation}
\label{SolASERL}
v(t,x)=c_1+c_2 x,
\end{equation}
and for the diffusion-wave regime $\alpha \in (1,2)$ has a solution 
\begin{equation}
\label{SolADWERL}
v(t,x)=c_1+c_2 x+(c_3+c_4 x) t.
\end{equation}
Here $c_i \ (i=1,2,3,4)$ are arbitrary constants. 

If Eq.~(\ref{TFDE}) with the Caputo fractional derivative is considered, then the right-sided Riemann-Liouville fractional derivative is used in the corresponding adjoint equation (\ref{AE}). This adjoint equation for the subdiffusion regime $\alpha \in (0,1)$ has a solution 
\begin{equation}
\label{SolASEC}
v(t,x)=(T-t)^{\alpha-1}(c_1 +c_2 x),
\end{equation}
and for the diffusion-wave regime $\alpha \in (1,2)$ has a solution 
\begin{equation}
\label{SolADWEC}
v(t,x)=(T-t)^{\alpha-2}\left[c_1+c_3 x + (T-t)(c_2+c_4 x)\right].
\end{equation}
Contrary to the solutions (\ref{SolASERL}) and (\ref{SolADWERL}), the solutions (\ref{SolASEC}) and (\ref{SolADWEC}) depend on the right time boundary $T$.

Because all presented solutions of the adjoint equation (\ref{AE}) are valid for any solution $u(t,x)$ to Eq.~(\ref{TFDE}), one can declare that nonlinear time-fractional subdiffusion and diffusion-wave equations with the Riemann-Liouville and Caputo fractional derivatives are nonlinearly self-adjoint. Therefore, the solutions (\ref{SolASERL}) -- (\ref{SolADWEC}) can be substituted into the formal Lagrangian (\ref{FL}) which then can be used for constructing conservation laws. 

%%%%%%%%%%%%%%
\subsection{Conservation laws for nonlinear TFDE with the Riemann-Liouville fractional derivative}
%%%%%%%%%%%%%%

Using the Noether operators (\ref{NOTRL}) and (\ref{NOX}), the symmetries (\ref{X_Abtr_k})--(\ref{X_k_u_43}), and the formal Lagrangian (\ref{FL}) with the function $v(t,x)$ given by (\ref{SolASERL}), only one new conserved vector has been found for Eq.~(\ref{TFDE}) with the Riemann-Liouville time-fractional derivative of order $\alpha \in (0,1)$. This conserved vector has the components
\begin{equation}
\label{CVSERL}
C^t=x \, {}^{}_0I^{1-\alpha}_t u, \ \ \ C^x=K(u)-x k(u) u_x,
\end{equation}
where $K(u)$ is an arbitrary function such that $K'(u)=k(u)$. Note that operator $X_1$ produces the trivial conserved vector for the constant $c_1$ from (\ref{SolASERL}), and the conserved vector (\ref{TCVRL}) for the constant $c_2$. Operators $X_2$ and $X_3$ give (\ref{TCVRL}) for the constant $c_1$, and (\ref{CVSERL}) for the constant $c_2$. Operator $X^{(1)}_4$ gives (\ref{CVSERL}) for the constant $c_1$ and the trivial conserved vector for the constant $c_2$.

In the case of $k(u)=u^{\frac{2\alpha}{1-\alpha}}$ using the symmetry (\ref{X_k_u_a}), two new conservation laws have been found. Corresponding conserved vectors have the components
\begin{equation}
\label{CVSERL1}
C^t=t \, {}^{}_0I^{1-\alpha}_t u - {}^{}_0I^{2-\alpha}_t u, \ \ \ C^x=- t u^{\frac{2\alpha}{1-\alpha}} u_x
\end{equation}
and
\begin{align}
\label{CVSERL2}
& C^t=x \left(t \, {}^{}_0I^{1-\alpha}_t u -  \,  {}^{}_0I^{2-\alpha}_t u \right), \nonumber \\ 
& C^x=t u^{\frac{2\alpha}{1-\alpha}} \left( \frac{1-\alpha}{1+\alpha} u - x  u_x \right).
\end{align}
The conserved vector (\ref{CVSERL1}) corresponds to the constant $c_1$, and conserved vector (\ref{CVSERL2}) corresponds to the constant $c_2$ from (\ref{SolASERL}). 

For the nonlinear diffusion-wave equation with the Riemann-Liouville time-fractional derivative, five new conservation laws have been found. The components of corresponding conserved vectors are presented in Table~\ref{Table_CVDWERL}, where conserved vector number 1 is the known conserved vector (\ref{TCVRL}). As previously, in this table $K(u)$ is an arbitrary function such that $K'(u)=k(u)$.

\begin{table}
\caption{Conserved vectors for the diffusion-wave equation with the Riemann-Liouville fractional derivative}
\label{Table_CVDWERL}
\begin{tabular}{ll}
\hline\noalign{\smallskip}
No & Components of the conserved vectors \\
\noalign{\smallskip}
\hline\noalign{\smallskip}
1. & $C^t={}^{}_0D^{\alpha-1}_t u$ \\
   & $C^x=-k(u)u_x$ \\ \hline\noalign{\smallskip}
2. & $C^t=t \, {}^{}_0D^{\alpha-1}_t u-{}^{}_0I^{2-\alpha}_t u$ \\
   & $C^x=-tk(u)u_x$ \\ \hline\noalign{\smallskip}
3. & $C^t=x \, {}^{}_0D^{\alpha-1}_t u$ \\ 
   & $C^x=K(u)-xk(u)u_x$ \\ \hline\noalign{\smallskip}
4. & $C^t=tx \, {}^{}_0D^{\alpha-1}_t u- x \, {}^{}_0I^{2-\alpha}_t u$ \\
   & $C^x=tK(u)-txk(u)u_x$ \\ \hline\noalign{\smallskip}
5. & $C^t=t^2 \, {}^{}_0D^{\alpha-1}_t u -2t \, {}^{}_0I^{2-\alpha}_t u +2 {}^{}_0I^{3-\alpha}_tu$ \\
   & $C^x=-t^2 k(u) u_x$ \\ \hline\noalign{\smallskip}
6. & $C^t=t^2 x \, {}^{}_0D^{\alpha-1}_t u -2tx \, {}^{}_0I^{2-\alpha}_t u +2x \, {}^{}_0I^{2-\alpha}_t u$ \\
   & $C^x=t^2K(u)-t^2xk(u)u_x$\\ 
\noalign{\smallskip}\hline
\end{tabular}
\end{table}

The correspondence between the symmetries (\ref{X_Abtr_k})--(\ref{X_k_u_a}), the constants $c_i \ (i=1,2,3,4)$ from (\ref{SolADWERL}), and the conserved vectors numbers from Table~\ref{Table_CVDWERL} is established by Table~\ref{Table_XCCVRL}. In this table index 0 corresponds to the trivial conserved vectors.

It is interesting to note that contrary to the linear case, the conserved vectors for the nonlinear TFDE (\ref{TFDE}) with the Riemann-Liouville time-fractional derivative do not involve the integral (\ref{J}). Moreover, the obtained conserved vectors for the nonlinear TFDE do not depend on the right time boundary $T$. 

\begin{table}
\caption{The correspondence between symmetries and conserved vectors numbers for the diffusion-wave equation with the Riemann-Liouville fractional derivative}
\label{Table_XCCVRL}
\begin{tabular}{cccccc}
\hline\noalign{\smallskip}
& $X_1$ & $X_2$ & $X^{(1)}_3$ & $X_4^{(1)}$ & $X_4^{(2)}$\\
\noalign{\smallskip}\hline\noalign{\smallskip}
$c_1$ & 0 & 1 & 1 & 3 & 2 \\
$c_2$ & 1 & 3 & 3 & 0 & 4 \\
$c_3$ & 0 & 2 & 2 & 4 & 5 \\
$c_4$ & 2 & 4 & 4 & 0 & 6 \\
\noalign{\smallskip}\hline
\end{tabular}
\end{table}

\subsection{Conservation laws for nonlinear TFDE with the Caputo fractional derivative}

Using the Noether operators (\ref{NOTC}) and (\ref{NOX}), the symmetries (\ref{X_Abtr_k})--(\ref{X_k_u_43}), (\ref{X_k_e_u}), and the formal Lagrangian (\ref{FL}) with the function $v(t,x)$ given by (\ref{SolASEC}), four new conservation laws have been found for the subdiffusion equation (\ref{TFDE}) with the Caputo fractional derivative. The corresponding conserved vectors are presented in Table~\ref{Table_CVSEC}, where function $\Phi(t)$ is defined as
$$
\Phi(t)=\frac{1}{\alpha \Gamma(1-\alpha)} \left(1-\frac{t}{T}\right)^{\alpha} {_2}F_1\left(\alpha,\alpha; \alpha+1; 1-\frac{t}{T} \right).
$$
Here ${_2}F_1(,;;)$ is the Gauss hypergeometric function.

\begin{table}
\caption{Conserved vectors for the subdiffusion equation with the Caputo fractional derivative}
\label{Table_CVSEC}
\begin{tabular}{ll}
\hline\noalign{\smallskip}
No & Components of the conserved vectors \\
\noalign{\smallskip}\hline\noalign{\smallskip}
1. & $C^t=u(0,x) \Phi(t)+(T-t)^{\alpha} \, {}^{}_0I^{1-\alpha}_t \left(\frac{u}{T-t} \right)$ \\
   & $C^x=-(T-t)^{\alpha-1}k(u)u_x$ \\ \hline\noalign{\smallskip}
2. & $C^t=(T-t)^{\alpha-1} \, {}^{}_0I^{2-\alpha}_t \left(\frac{u_t}{T-t} \right)$ \\ 
   & $C^x=-(T-t)^{\alpha-2}k(u)u_x$ \\ \hline\noalign{\smallskip}
3. & $C^t=x u(0,x) \Phi(t)+x(T-t)^{\alpha} \, {}^{}_0I^{1-\alpha}_t \left(\frac{u}{T-t} \right)$ \\
   & $C^x=(T-t)^{\alpha-1}\left[K(u)-xk(u)u_x\right]$ \\ \hline\noalign{\smallskip}
4. & $C^t=x(T-t)^{\alpha-1} \, {}^{}_0I^{2-\alpha}_t \left(\frac{u_t}{T-t} \right)$ \\
   & $C^x=(T-t)^{\alpha-2}\left[K(u)-xk(u)u_x\right]$ \\ 
\noalign{\smallskip}\hline
\end{tabular}
\end{table}

The correspondence between the symmetries (\ref{X_Abtr_k})--(\ref{X_k_u_43}), (\ref{X_k_e_u}), the constants $c_1$ and $c_2$ from (\ref{SolADWERL}), and the  conserved vectors numbers from Table~\ref{Table_CVSEC} is established by Table~\ref{Table_XCCVC}. Thus, symmetry $X_2$ produces all four conservation laws. The trivial conserved vectors are produced by the operator $X_1$ for the constant $c_1$ and by the operator $X_4^{(1)}$ for the constant $c_2$. In Table~\ref{Table_XCCVC}, the trivial conserved vectors are denoted by 0. 

\begin{table}
\caption{The correspondence between symmetries and conserved vectors numbers for the subdiffusion equation with the Caputo fractional derivative}
\label{Table_XCCVC}
\begin{tabular}{cccccc}
\hline\noalign{\smallskip}
& $X_1$ & $X_2$ & $X^{(1)}_3$ & $X^{(2)_3}$ & $X_4^{(1)}$ \\
\noalign{\smallskip}\hline\noalign{\smallskip}
$c_1$ & 0 & 1, 2 & 1 & 1 & 3 \\
$c_2$ & 1 & 3, 4 & 3 & 3 & 0 \\
\noalign{\smallskip}\hline
\end{tabular}
\end{table}

For the nonlinear diffusion-wave equation with the Caputo time-fractional derivative, six new conserved vectors have been found. The components of these vectors are presented in Table~\ref{Table_CVDWEC}. As previously, in this table $K'(u)=k(u)$. Also, the following notations are used in Table~\ref{Table_CVDWEC} (here $1 < \alpha <2)$: 
\begin{multline*}
\Phi(t)=\frac{1}{(\alpha-1)\Gamma(2-\alpha)} \left(1-\frac{t}{T}\right)^{\alpha-1} \\
 				\times {_2}F_1\left(\alpha-1,\alpha-1;\alpha;1-\frac{t}{T} \right),
\end{multline*}
\begin{multline*}
\Psi(t)=\frac{1}{\alpha\Gamma(2-\alpha)} \left(1-\frac{t}{T}\right)^{\alpha} \\ 
				\times {_2}F_1\left(\alpha-1,\alpha;\alpha+1;1-\frac{t}{T} \right),
\end{multline*}
\begin{multline*}
\left( {}^{F}_0I^{2-\alpha}_t f \right)(t) = \frac{1}{\Gamma(2-\alpha)} \\
				\times \int_0^t \frac{f(\tau)}{(t-\tau)^{\alpha-1}} {_2}F_1\left(1,1;2-\alpha; \frac{t-\tau}{T-\tau} \right) d \tau.
\end{multline*}

\begin{table}
\caption{Conserved vectors for the diffusion-wave equation with the Caputo time-fractional derivative}
\label{Table_CVDWEC}
\begin{tabular}{ll}
\hline\noalign{\smallskip}
No & Components of the conserved vectors \\
\noalign{\smallskip}\hline\noalign{\smallskip}
1. & $C^t=(T-t)^{\alpha-2} \, {}^{}_0I^{3-\alpha}_t \left(\frac{u_{tt}}{T-t} \right)$ \\
   & $C^x=-(T-t)^{\alpha-3} \, k(u)u_x$ \\ \hline\noalign{\smallskip}
2. & $C^t=\Phi(t) u_t(0,x) + (T-t)^{\alpha-1} \, {}^{}_0I^{2-\alpha}_t \left(\frac{u_t}{T-t} \right)$ \\
   & $C^x=-(T-t)^{\alpha-2} \, k(u)u_x$ \\ \hline\noalign{\smallskip}
3. & $C^t=\Psi(t) u_t(0,x) + (T-t)^{\alpha} \, {}^{F}_0I^{2-\alpha}_t \left(\frac{u_t}{T-t} \right)$ \\
   & $C^x=-(T-t)^{\alpha-1} \, k(u)u_x$ \\ \hline\noalign{\smallskip}
4. & $C^t=x(T-t)^{\alpha-2} \, {}^{}_0I^{3-\alpha}_t \left(\frac{u_{tt}}{T-t} \right)$ \\
   & $C^x=(T-t)^{\alpha-3} \left(K(u)-x k(u)u_x \right)$ \\ \hline\noalign{\smallskip}
5. & $C^t=x\Phi(t) u_t(0,x) + x (T-t)^{\alpha-1} \, {}^{}_0I^{2-\alpha}_t \left(\frac{u_t}{T-t} \right)$ \\
   & $C^x=(T-t)^{\alpha-2} \left(K(u)-x k(u)u_x \right)$ \\ \hline\noalign{\smallskip}
6. & $C^t=x\Psi(t) u_t(0,x) + x(T-t)^{\alpha} \, {}^{F}_0I^{2-\alpha}_t \left(\frac{u_t}{T-t} \right)$ \\
   & $C^x=(T-t)^{\alpha-1} \left(K(u)-x k(u)u_x \right)$ \\ 
\noalign{\smallskip}\hline
\end{tabular}
\end{table}

The correspondence between the symmetries (\ref{X_Abtr_k})--(\ref{X_k_u_43}), (\ref{X_k_e_u}), the constants $c_i$ ($i=1,2,3,4$) from (\ref{SolADWEC}), and the   conserved vectors numbers from Table~\ref{Table_CVDWEC} is established by Table~\ref{Table_XCCVC1}. Thus, the symmetry $X_2$ produces all six conservation laws. The trivial conserved vectors have been obtained by the operator $X_1$ for the constants $c_1$ and $c_2$, and by the operator $X^{(1)}_4$ for the constants $c_3$ and $c_4$. In Table~\ref{Table_XCCVC1}, the trivial conserved vectors are denoted by 0.

\begin{table}
\caption{The correspondence between symmetries and conserved vectors numbers for the diffusion-wave equation with the Caputo fractional derivative}
\label{Table_XCCVC1}
\begin{tabular}{cccccc}
\hline\noalign{\smallskip}
& $X_1$ & $X_2$ & $X^{(1)}_3$ & $X_3^{(2)}$ & $X_4^{(1)}$\\
\noalign{\smallskip}\hline\noalign{\smallskip}
$c_1$ & 0 & 1,2 & 2 & 2 & 5 \\
$c_2$ & 0 & 2,3 & 3 & 3 & 6 \\
$c_3$ & 2 & 4,5 & 5 & 5 & 0 \\
$c_4$ & 3 & 5,6 & 6 & 6 & 0 \\
\noalign{\smallskip}\hline
\end{tabular}
\end{table}

Finally, one additional remark should be done. If $u_t(0,x)=0$, then the operator $X_4^{(2)}$ given by (\ref{X_k_u_a}) is admitted by Eq.~(\ref{TFDE}) with the Caputo time-fractional derivative of order $\alpha \in (1,2)$ and $k(u)=u^{\frac{2\alpha}{1-\alpha}}$. So, this operator can be considered as a conditional symmetry for this equation. This operator produces all six conserved vectors from Table~\ref{Table_CVDWEC}: vectors 1, 2, 3 for the constant $c_1$, vectors 2, 3 for the constant $c_2$, vectors 4, 5, 6 for the constant $c_3$, and vectors 5, 6 for the constant $c_4$.

%----------------------------------
\section{Conclusion}
\label{sec:conc}
%----------------------------------

The approach described in this paper allows one to construct conservation laws for fractional differential equations with the Riemann-Liouville and Caputo fractional derivatives of order $\alpha \in (0,2)$ that do not have Lagrangians in classical sense. Moreover, this approach can be extended to the fractional differential equations with another types of fractional derivatives, such as Erdelyi-Kober derivative, Hadamard derivative, Riesz derivative, etc.

\section*{Acknowledgements}
This work was supported by the grant of the Ministry of Education and Science of the Russian Federation (contract No.~11.G34.31.0042 with Ufa State Aviation Technical University and leading scientist Professor N.~H.~Ib\-ra\-gi\-mov). The author is also grateful to Professor Rafail K. Gazizov for helpful discussion of the manuscript.

\end{document}